\documentclass[12pt, twocolumn]{iopart}

\newcommand{\CCO}{CuCrO$_2$}
\newcommand{\CFO}{CuFeO$_2$}
\newcommand{\KCO}{K$_2$Cr$_2$O$_7$}

\usepackage{graphicx}
\usepackage{dcolumn}
\usepackage{bm}
\usepackage{mathrsfs}
\usepackage{appendix}
\usepackage{bbm}
\usepackage{color}
\usepackage{cite}
\usepackage{textcomp}
\usepackage{iopams}
\newcommand{\nsub}[1]{_{\rm{#1}}}

\newcommand{\grad}{\ensuremath{^\circ}}

\newcommand{\mub}{$\mu\nsub{B}$}

\begin{document}

\title{Magnetic structure of \CCO: a single crystal neutron diffraction study}%

\author{M. Frontzek$^1$, G. Ehlers$^1$, A. Podlesnyak$^1$, H. Cao$^1$, M. Matsuda$^1$, O. Zaharko$^2$, N. Aliouane$^2$, S. Barilo$^3$, S.V.
Shiryaev$^3$}

\address{$^1$Neutron Scattering Science Division, Oak Ridge National
Laboratory, Oak Ridge, TN 37831, USA}
\address{$^2$Laboratory for Neutron Scattering, Paul Scherrer Institute,
CH-5232 Villigen, Switzerland}
\address{$^3$Institute of Solid State and Semiconductor Physics, Minsk
220 072, Belarus}

\ead{frontzekmd@ornl.gov}

\begin{abstract}
This paper presents results of a recent study of multiferroic \CCO\
by means of single crystal neutron diffraction. This system has two
close magnetic phase transitions at $T\nsub{N1}=24.2$~K and
$T\nsub{N2}=23.6$~K. The low temperature magnetic structure below
$T\nsub{N2}$ is unambiguously determined to be a fully 3-dimensional
proper screw. Between $T\nsub{N1}$ and $T\nsub{N2}$
antiferromagnetic order is found that is essentially 2-dimensional.
In this narrow temperature range, magnetic near neighbor
correlations are still long range in the ($H,K$) plane, whereas
nearest neighbors along the $L$-direction are uncorrelated. Thus,
the multiferroic state is realized only in the low-temperature
3-dimensional state and not in the 2-dimensional state.

\end{abstract}


\maketitle

\section{Introduction}

\label{Introduction}

Multiferroic materials have become of interest for their unusual
low-temperature properties in general, and in particular for the
observation that one can affect their magnetic structure through an
electric field and their electric polarization through a magnetic
field. The most promising candidates for a controllable multiferroic
have been found among the materials with inherent geometric magnetic
frustration~\cite{Che07}.

The magnetic properties of the delafossite \CCO\ have received
detailed interest since the discovery of its multiferroic
behaviour~\cite{Kim08,Kim09,Yama10}. This system, which crystallizes
in the rhombohedral $R\overline{3}m$ space group, is a multiferroic
compound due to its apparent strong coupling of spin and charge. In
contrast to other multiferroic compounds \CCO\ shows a spontaneous
electric polarization upon antiferromagnetic ordering without an
accompanying structural phase transition, although a slight in-plane
lattice distortion can be measured~\cite{Kim09b}. Further, \CCO\ is
a rare example of a system whose magnetoelectric properties are
tunable by both an electric and a magnetic
field~\cite{Kim09,Soda09}. Another particular property is that the
multiferroic state is already entered in zero magnetic field as
opposed to isostructural \CFO~\cite{Pet00,Kim06,Ye06}.

Several studies of the magnetic structure by neutron diffraction
techniques with powder~\cite{Kad90,Poi09} and single
crystal~\cite{Soda09,Soda10} samples have been reported. The powder
experiments found an incommensurate magnetic propagation vector
$\bm\tau=(0.329,0.329,0)$ based on the position of the
$(0,0,0)+\bm\tau$ reflection, and a broadening of some of the
magnetic Bragg peaks which was attributed to short range
correlations along the $L$-direction, either in the form of quasi
2-dimensionality~\cite{Kad90} or stacking faults~\cite{Poi09}. The
measured powder intensities were best described with a proper-screw
magnetic structure, although other magnetic structure models could
not be unambiguously excluded. The single crystal experiments, on
the other hand, identified the structure as an incommensurate proper
screw but the magnetic moments were not refined, and, most
importantly, measurements were only reported for ($H,K,0$)
reflections~\cite{Soda09,Soda10}. Thus the question of possible
ordering along the $L$-direction was not considered. In this
contribution we will fill these gaps and arrive at the following
main conclusions. The low-temperature magnetic structure below
$T\nsub{N2}$\footnote{opposite from \cite{Kim09} we define
$T\nsub{N2} < T\nsub{N1}$} is truly 3-dimensional with long range
correlations along the $L$-direction. We confirm the incommensurate
proper screw magnetic spiral propagating along the [1,1,0]
direction, and find that it has an elliptical envelope with a
modulation between $\bf{M}_{[1\overline{1}0]}$= 2.2(2)~\mub~and
$\bf{M}_{[001]}$= 2.8(2)~\mub. Furthermore we characterize the spin
structure in the intermediate temperature phase between $T\nsub{N1}$
and $T\nsub{N2}$ (which is not multiferroic) as 2-dimensional with a
total lack of near neighbor magnetic correlations along the
$L$-direction.

\section{Experimental}
\label{Experimental}

Single crystal samples were prepared in two ways. Samples up to
6~mm$^3$ in size were grown in a platinum crucible from the high
temperature solution technique based on the thermal decomposition of
\KCO\ at 860\grad{C} in the presence of CuO~\cite{Crot96}. The
mixture of \KCO\ (73 mol\%) and CuO (27 mol\%) was placed into a
high density aluminum crucible and heated quickly up to 850\grad{C}.
Then the temperature was lowered at a rate of 0.5$^\circ$C/h during
a period of up to four days before the crucible was quenched to room
temperature. This resulted in single crystalline black hexagonal
platelets of up to 6 mm$^3$ in volume. Larger \CCO\ crystals (up to
60 mm$^3$) were grown with a flux melting technique based on
Bi$_2$O$_3$ solvent at a temperature between 940\grad{C} and
1250\grad{C}, under conditions of repeating abrupt rise of
temperature of 10-15\grad{C} with subsequent cooling of
0.5-1.5\grad{C}/h.

The trigonal crystal structure (space group $R\bar{3}m$) with room
temperature lattice parameters $a=2.976$~{\AA} and
$c=17.110$~{\AA}~\cite{Crot96b} was confirmed by X-ray powder
analysis of crushed crystals. Further characterization with respect
to their magnetic properties was done using a SQUID-magnetometer.
The obtained susceptibility curves were similar to data published
previously~\cite{Kim08,Poi09,Kim09,Soda10}. Identifying the same
characteristic points in the susceptibility data as Kimura
\etal~\cite{Kim08} the same two characteristic phase transition
temperatures, $T\nsub{N1}=24.2$~K and $T\nsub{N2}=23.6$~K, were
obtained for our samples.


\begin{figure}[b]
\includegraphics[width=8.5cm]{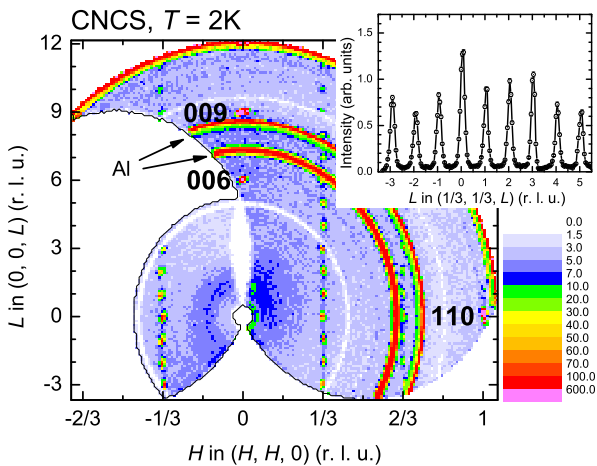}
\caption{(colour online) Reciprocal $HHL$ map of \CCO\ measured on
CNCS at $T=2$~K. The nuclear (110), (006) and (009) Bragg peaks as
well as the polycrystalline rings from the Al sample mount are
marked. The unmarked reflections are magnetic. The inset in the
upper right corner shows a section along $(1/3,1/3,L)$.}
\label{CNCSdiff}
\end{figure}

Neutron single crystal diffraction experiments were performed on the
TriCS four circle diffractometer at the SINQ facility, PSI,
Switzerland, on the HB-1 triple axis spectrometer and the HB-3A four
circle diffractometer (FCD) at HFIR at Oak Ridge National Laboratory
(ORNL), and on the Cold Neutron Chopper Spectrometer (CNCS) at the
Spallation Neutron Source (SNS) in Oak Ridge. The TriCS experiment
used a single crystal of dimensions 3.5 mm $\times$ 3 mm $\times$ 1
mm mounted in two different orientations: with the [00$L$] direction
(hexagonal setting) along the $\phi$-shaft of the instrument
(orientation 1) and with the [$\overline{H}H0$] direction
(orientation 2) along the $\phi$-shaft. Orientation 1 gives good
resolution in the $H,K$ planes ($L$=const), while orientation 2
yields good resolution in the planes orthogonal to
[$\overline{H}H0$]. For data collection in a broad $\bf{Q}$ range a
neutron wavelength of $\lambda=1.178$~{\AA} (Ge-311 monochromator
with vertical focusing \cite{Sch00}) was used without collimation
between the sample and the $^3$He single tube detector. For
collecting maps and scans with high resolution a neutron wavelength
of  2.32~{\AA} was used (PG-002 monochromator, PG filter) with a 20'
collimation in front of the detector. The experiments on HB-3A FCD
at HFIR used a single crystal of similar size, 3.2 mm $\times$ 2.5
mm $\times$ 1.0 mm, from the same batch. The crystal was mounted in
the orientation 2. The neutron wavelength was $\lambda=1.536$~{\AA}
(Si-220 monochromator in high resolution mode (bending
150)~\cite{Cha11}) with a 8 mm pinhole mask in front of the
detector. Maps collected on the TriCS diffractometer used a 40 step
grid with a stepsize of 0.001 reciprocal lattice units (r. l. u.).
On HB-3A a 30 step grid with 0.002 r. l. u. step size was used. In
order to avoid confusion, the $HK$-values of the scans have been
transformed to an orthogonal system using the $HH$ direction as $x$
and the $\overline{H}H$ direction as $y$. Additional data were taken
with elastic measurements on the HB-1 spectrometer at HFIR and on
CNCS at the SNS \cite{CNCS}. These experiments were performed with
the intent to measure the magnetic excitation spectrum and therefore
the sample consisted of 10 co-aligned single crystals in ($HHL$)
scattering geometry mounted on an aluminium sheet. On HB-1 the
neutron wavelength of 2.46~{\AA} from a PG-002 monochromator was
used. The collimation was 48'-60'-60'-240' with additional PG
filters in the incident beam. The incident wavelength at CNCS was
2.60~{\AA}.

Two data sets of nuclear Bragg reflections (40 and 65 reflections
for the orientations 1 and 2, respectively) were collected on TriCS
for a scale factor determination at 300 K and at 5 K with
$\lambda=1.178$~{\AA}. These measurements proved the absence of
reverse/obverse twinning~\cite{She02} and confirmed the published
nuclear structure~\cite{Crot96b}.

\begin{figure}[b]
\includegraphics[width=8.0cm]{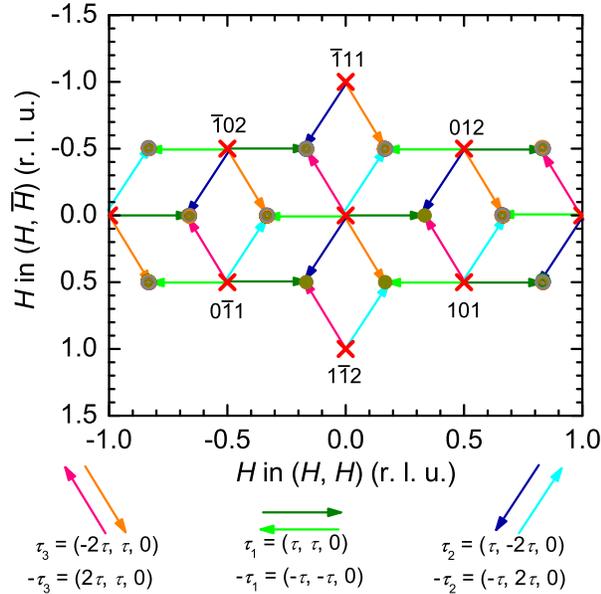}
\caption{(Colour online) $HK$-projection of the magnetic reflections
(circles) and their origin. The crosses mark the position of the
nuclear reflections. The star of the propagation vector $\tau$ is
indicated by the different coloured arrows.} \label{Scheme}
\end{figure}

\section{Results}
\label{Results}

A diffraction map measured at $T=2$~K on CNCS is shown in
figure~\ref{CNCSdiff}. With this pattern it is clear that the
magnetic order is truly 3-dimensional, in fact all measurements find
the full width at half maximum (FWHM) in the $L$-direction to be
resolution limited (insert in ~\ref{CNCSdiff}). The correlation
length is therefore larger than at least several hundred ~{\AA}.
This result clearly disagrees with the conclusions from the powder
diffraction measurements~\cite{Kad90,Poi09} (which were suggesting
short range order along the $L$-direction).

\begin{figure}[t]
\includegraphics[width=8.5cm]{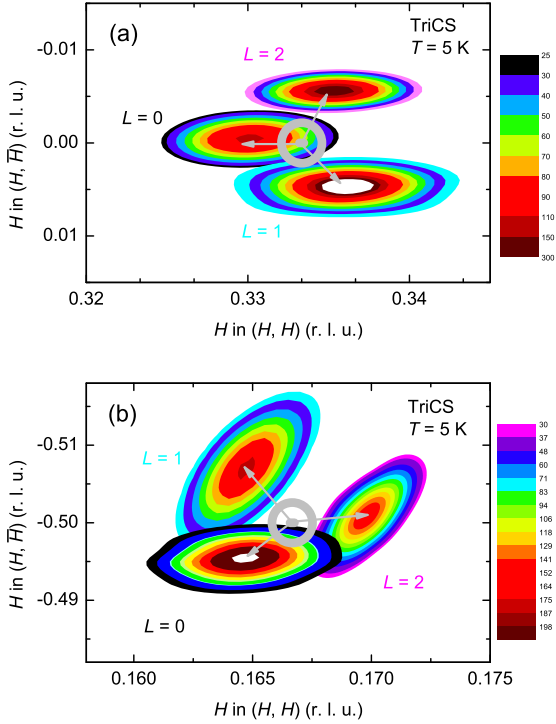}
\caption{(Colour online) Upper panel: $H,K$-projection of the
magnetic reflections around the $(1/3,1/3,L)$ position (grey point)
as measured on TriCS. Lower panel: magnetic reflections around the
$(-1/3,2/3,L)$ position} \label{Fig3Combined}
\end{figure}

To identify the fundamental magnetic propagation vector, a detailed
mapping of the magnetic reflections in the basal plane(s) has been
performed at $T=5$~K for both the TriCS and HB-3A datasets.
Figure~\ref{Scheme} gives an overview over the measured magnetic
reflections and the related nuclear reflections to which they are
satellite. The results are shown for the magnetic reflections with
$L=0,1,2$ around $(1/3,1/3,L)$ and around $(-1/3,2/3,L)$ in
figure~\ref{Fig3Combined}. The figure shows the projection of the
three observed reflections in the $H,K$-plane. The magnetic
reflections are located around the commensurate 1/3 position which
is also shown.

Based on our measurements, the magnetic structure of \CCO\ can be
satisfactorily described  with a pair of incommensurate propagation
vectors ($\bm{\tau_1}$,-$\bm{\tau_1}$) with
$\bm{\tau_1}=(\tau,\tau,0)$ and $\tau=0.3298(1)$. By symmetry, the
star of the propagation vector consists of
$\bm{\tau_1}=(\tau,\tau,0)$, $\bm{\tau_2}=(\tau,-2\tau,0)$ and
$\bm{\tau_3}=(-2\tau,\tau,0)$.

To confirm the magnetic structure a refinement of the two datasets
measured at TriCS, collected with $\lambda$=1.178~\AA~(140
reflections) and $\lambda$=2.32~\AA~(197 reflections) has been
performed using FullProf \cite{Full}. Three magnetic domains due to
the 3-fold axis have been taken into account. The following four
models for the arrangement of magnetic moments were considered:
\begin{itemize}
\item{amplitude modulated structure derived from the ``Proper Helix"}
\item{``Cycloid 1" with moments rotating in the [110]-[001]-plane}
\item{``Cycloid 2" with moments rotating in the [110]-[1$\overline{1}$0]-plane}
\item{the ``Proper Helix" with moments rotating in the [1$\overline{1}$0]-[001]-plane as in \cite{Soda10}}
\end{itemize}
The possibility of an elliptical envelope was allowed for the three
last cases. Generally the magnetic moment $\bf{M}$ of the Cr$^{3+}$
ion in the $l$th unit cell can be described by the generalized helix
presentation,
\begin{displaymath}
\bf{M}_l=\bf{M}_{1}\cos(\bm\tau \cdot {\bf{r}}_l+\psi) +
\bf{M}_{2}\sin(\bm\tau \cdot {\bf{r}}_l+\psi) {\;},
\end{displaymath}
where $\bf{M}_{1}$ and $\bf{M}_{2}$ are orthogonal base vectors
determining the magnitude and direction of the generalized helix and
$\psi$ is the phase.

The results for our refinement are listed in table 1~\ref{table1}.
The cycloid models fit the experimental data poorly since they
predict small intensities for the magnetic satellites along the
wavevectors (i.e $\bm{\tau_1}$,$\bm{\tau_2}$,$\bm{\tau_3}$).
\begin{table*}[h]
\begin{center}
\caption{Reliability factors for the magnetic structure at $T=5$~K
measured with $\lambda$=1.178~\AA~. Four different proto-type
magnetic structures have been considered: Amplitude modulated,
Cycloid 1 ($ab$-cycloid), Cycloid 2 ($110$-$c$-cycloid) and Proper
helix. The refined parameters were the magnetic moment and the
domain population.} \label{table1}

 \scalebox{0.7}{
\begin{tabular}{ccccc}
 \hline
 \multicolumn{5}{c}{Refined set of 140 magnetic reflections measured with 1.178\AA}  \\ \hline
                  & Ampl. Mod. &     ~Cycloid 1&  Cycloid 2 &   Proper helix       \\ \hline
       RF$^{2}$   &      35.0       &      43.1         &  35.1     &    28.0       \\
       RF$^{2}$w  &      38.1       &      44.7        & 38.1       &      30.6     \\
       RF         &      20.8       &      24.0       & 20.8       &        13.7     \\
       \hline
      M$_1$($\mu_B$)&  [1$\overline{1}$0]  -0.1(6)    & [110] 1(1)       &   [110] 0(8)     & [1$\overline{1}$0]   2.2(2)  \\
      M$_2$($\mu_B$)&    [001] ~3.3(2)                & ~~[1$\overline{1}$0] 3.2(4)       &~   [001] 3.4(4)    & [001] 2.8(2)  \\ \hline
      Domain population (\%)&    ~39(7)~:~36(6)~:~25(7)~    &     ~39(9)~:~40(9)~:~22(9)~   &    ~39(8)~:~36(8)~:~25(9)~  &    ~37(2)~:~37(2)~:~26(3)~      \\ \hline
\end{tabular}}
\end{center}
\end{table*}

\begin{figure}[t]
\includegraphics[width=8.5cm]{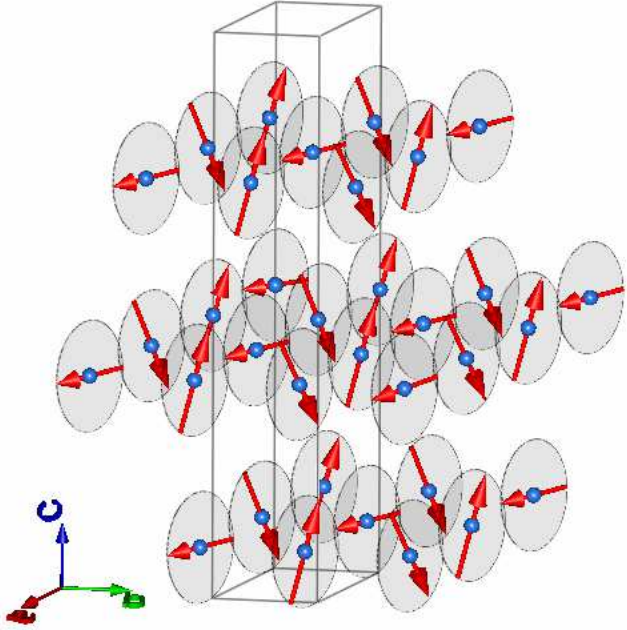}
\caption{(Colour online) Real space representation of the magnetic
structure of \CCO. The chemical unit cell is outlined.  }
\label{Magstru}
\end{figure}

In reality these intensities are large. This indicates that the
magnetic moment is predominantly orthogonal to the wavevector. The
amplitude modulated model agrees slightly better with the data, but
the agreement factors are larger than the ones obtained with the
``Proper Helix" model. Therefore, we agree with the powder
diffraction results of Poienar \etal\cite{Poi09} and polarized data
of Soda \etal\cite{Soda10} that the magnetic structure in \CCO\ is a
proper helix. The statistically best fit is obtained for an
elliptical helix with the moments averaged over all domains
$\bf{M}_{[1\overline{1}0]}$= 2.2(2)~\mub~and $\bf{M}_{[001]}$=
2.8(2)~\mub. The data yields a different population for each domain.
For the helix the obtained population is
~37(2)\%~:~37(2)\%~:~26(3)\%~. In principle $\bm{\tau_1}$ and
-$\bm{\tau_1}$ are not equivalent wave vectors and they might
correspond to two different structures. Such situation is however
energetically not favored and we considered $\bm{\tau_1}$ and
-$\bm{\tau_1}$ as generators to two chiral domains of the same
structure. Figure \ref{Magstru} depicts the real space
representation of the ``Proper Helix" model.

\begin{figure}[t]
\includegraphics[width=8.5cm]{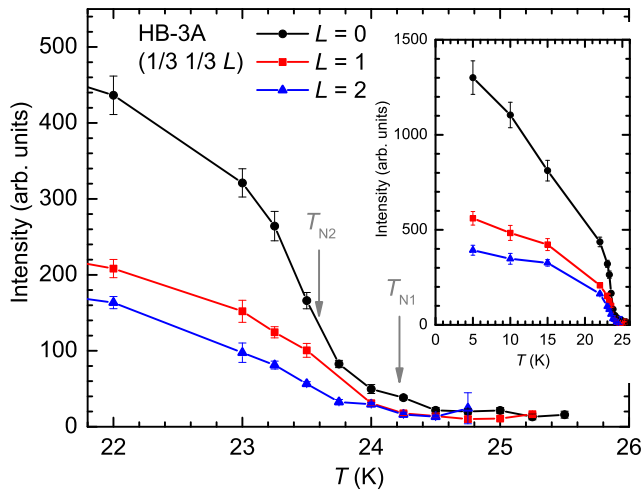}
\caption{(Colour online) Intensity vs. temperature plot for the
magnetic reflections around $(1/3,1/3,L)$ in the region from 22 to
26 K from measurements on HB-3A. The two transition temperatures
$T\nsub{N1}$ and $T\nsub{N2}$ are marked. The entire temperature
range is shown as inset.} \label{TdepPeaks}
\end{figure}

\begin{figure}[t]
\includegraphics[width=8.5cm]{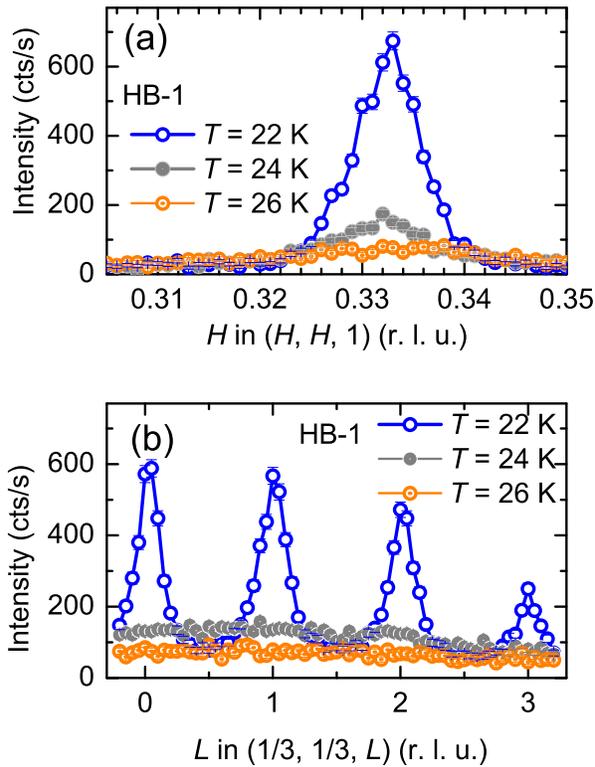}
\caption{(Colour online) Magnetic intensities at three temperatures
measured along the $HH$ direction (a) and along the $L$ direction
(b) on HB-1.} \label{DirDepPeaks}
\end{figure}

We next discuss the temperature dependence of the magnetic
structure. The magnetic intensity as shown in figure~\ref{TdepPeaks}
decreases strongly with temperature above $\sim{23}$~K and would be
in agreement with a critical temperature slightly below 24~K.
However, the magnetic intensity above 24~K is still significant.
Therefore the half-widths of the magnetic reflections were also
analyzed along different lattice directions.
Figure~\ref{DirDepPeaks} shows three measurements for both the $H,H$
and the $L$ direction, below $T\nsub{N2}$, between $T\nsub{N1}$ and
$T\nsub{N2}$, and above $T\nsub{N1}$, respectively. Measured along
the $H,H$-direction the magnetic Bragg peak loses roughly a factor
of four in intensity between 22~K and 24~K but the width changes
only slightly.
 At 26~K
the magnetic intensity is then close to the background and the
intensity broadly smeared as expected above (but close to) the
magnetic ordering temperature. Apparently $T\nsub{N2}$ is not a
critical temperature along this direction. In the lower panel of
figure~\ref{DirDepPeaks} measurements along the $L$-direction at the
same temperatures are shown for comparison. Here, no sign of a Bragg
peak can be seen in the intermediate temperature range, whereas
along $H,H$ there is still a peak, albeit small, at the same
temperature. Along the $L$-direction one finds {\em diffuse}
magnetic scattering between $T\nsub{N1}$ and $T\nsub{N2}$, well
above instrument background, indicative of a complete loss of the
magnetic near neighbor correlations along this direction. Since the
measurement uses energy analysis, the magnetic correlations can be
considered static.

\begin{figure}[t]
\includegraphics[width=8.5cm]{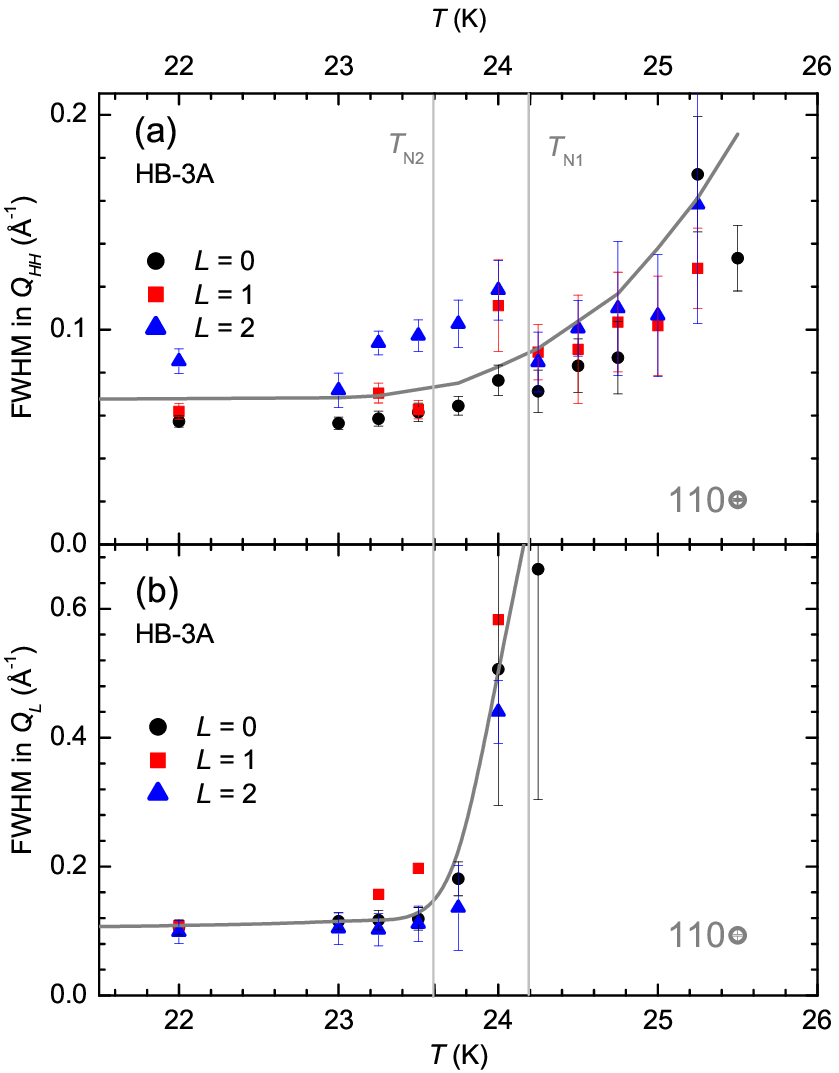}
\caption{(Colour online) Temperature dependence of the FWHM of the
magnetic satellites measured on HB-3A along (a) the $HH$ direction
and (b) along the $L$ direction in the temperature range from 22 K
to 26 K. The lines are guides to the eye. The gray points at 25.5 K
are the measured FWHM of the nuclear 110 reflection at room
temperature. The vertical gray lines mark the transition
temperatures $T\nsub{N1}$ and $T\nsub{N2}$. The width of the Bragg
peaks along $L$ diverges approx. 1~K below the temperature at which
it diverges along $HH$.} \label{TdepWidths}
\end{figure}

The picture becomes clear with an analysis of the temperature
dependence of the widths of the Bragg peaks. This is shown in
figure~\ref{TdepWidths}. The trend for the peak width along the $HH$
direction is shown in the upper panel of figure~\ref{TdepWidths}.
The width increases at increasing temperatures starting around 24~K.
Above 25~K it then becomes difficult to fit a  peak to the data. In
contrast, the width along the $L$-direction as shown in the lower
panel of the figure start to significantly increase already above
23.6~K. At 24~K the fitted width is already as large as the distance
between reciprocal lattice points.

To summarize, the picture that emerges is that \CCO\ enters a truly
2-dimensional ordered state between $T\nsub{N1}$ and $T\nsub{N2}$,
with well developed long range correlations in the $H,K$ plane but a
lack of correlation in the $L$-direction. Only below $T\nsub{N2}$  a
3-dimensional ordering is established.

\section{Discussion}
The new single crystal diffraction data and our analysis show that
the low temperature magnetic ordering of \CCO\ is fully
3-dimensional and can be described as a incommensurate proper helix
propagating in the [$H,H,0$] direction. Moments are in the
[$H,\overline{H},L$] plane. The envelope of the spiral is slightly
elliptic. The magnetic propagation vector $\bm\tau$ is very close to
the commensurate (1/3,~1/3,~0) value. The commensurate case would
correspond to the classic 120\grad\ magnetic ground state of the
2-dimensional triangular lattice. The cause of the
incommensurability of the propagation vector has been discussed
earlier, and two different explanations have been proposed: either
an in-plane lattice distortion \cite{Kim09b,Yama10} or interlayer
exchange interaction~\cite{Ras86,Ras87,Ras88,Kad95}. The latter is
in better agreement with the occurrence of 3-dimensional order,
although it had been shown that an interlayer exchange has to be
relatively small ($<$0.2meV)~\cite{Poi10}. The maximum ordered
moment in our model is $2.8(2)$~\mub~ which is close to the expected
$3$~\mub~ for the Cr$^{3+}$ ion and the reported values from the
powder measurements~ \cite{Kad90,Poi09}. An additional uncertainty
for the absolute moment value from the single crystal measurements
comes from the non-equal domain distribution. A cycloidal spin
structure can be excluded from our analysis, in accordance with the
observation of multiferroicity in the low temperature phase and its
explanation after Arima~\cite{Arima07,Soda09,Soda10}. The 3D
character of the low temperature ordering is in agreement with the
interpretation of the
earlier published specific heat data at the phase transition~\cite{Poi09}.\\
The multiferroic state of \CCO\ is entered without an apparent
structural symmetry reduction or lattice distortion. Within the
resolution of our experiments the upper limit for change of the
lattice constants at the phase transition is 0.001~{\AA}. However,
evidence for in-plane distortions have been observed in high
resolution X-ray diffraction measurements\cite{Kim09b}. This is in
agreement with small structural distortion observed in the related
\CFO\
system~\cite{Ye06}. \\
Why is \CCO\ not multiferroic between $T\nsub{N1}$ and $T\nsub{N2}$?
According to the Arima model \cite{Arima07} the in-plane proper
screw spiral will create a spontaneous polarization even without the
observed 3-dimensional order. The propagation vector in the narrow
phase is the same as in the multiferroic phase and it will be argued
in the following why the magnetic structure in the intermediate
phase is most likely an in-plane proper screw but without interlayer
order. In the light of the 2-dimensional nature this narrow phase
has been discussed as a possible collinear state~\cite{Kim08}.
However, it is unlikely that the narrow temperature phase has a
collinear structure. It has been shown theoretically \cite{Fis10}
that the collinear state is energetically not favored. Also, a
collinear state with three sublattices would have a net
ferromagnetic moment (which is not observed). A re-ordering from
collinear to helical state is expected to feature a large response
in the temperature dependent susceptibility (which is not observed).
Similar, an amplitude modulated magnetic structure on a three sub
lattice without a ferromagnetic net moment would undergo a
discontinues phase transition to a spiral state.

The same propagation vector in both phases with the absence of a net
ferromagnetic component and the continuous course of the
susceptibility through the transition at $T\nsub{N2}$ indicate that
spirals already form at $T\nsub{N1}$. The uncorrelated spirals then
order through ferromagnetic interlayer exchange to a full
3-dimensional order at $T\nsub{N2}$. The question why \CCO\ is not
multiferroic between $T\nsub{N1}$ and $T\nsub{N2}$ cannot be
answered with the results from the diffraction experiment alone
since in the picture of the uncorrelated spirals the spontaneous
polarization could be averaged to zero. A possible experiment to
clarify this question is neutron diffraction with an applied
electric field in this narrow phase.

\ack{We acknowledge the technical and scientific support from the
staff at SNS, HFIR, and PSI. This work was partly performed at SINQ,
Paul Scherrer Institute, Villigen, Switzerland. This research was
sponsored by the Division of Materials Sciences and Engineering of
the U. S. Department of Energy. Research at Oak Ridge National
Laboratory's Spallation Neutron Source was supported by the
Scientific User Facilities Division, Office of Basic Energy
Sciences, U. S. Department of Energy. The work in Minsk was
supported in part by Belarusian Fund for Basic Scientific Research,
grant No  F10R-154.}

\pagebreak


\end{document}